\documentclass[reprint]{revtex4-1}

\usepackage{graphicx} 
\usepackage{amsmath}
\usepackage{ amssymb }

\usepackage{color}

\usepackage{float} 
\usepackage{wrapfig} 

\usepackage[makeroom]{cancel} 
\usepackage{comment}

\usepackage{lipsum} 

\newcommand{\beq}{\begin{equation}}
\newcommand{\eeq}{\end{equation}}

\newcommand{\bvec}{\begin{pmatrix}}
\newcommand{\evec}{\end{pmatrix}}
\newcommand{\lp}{\left(}
\newcommand{\rp}{\right)}
\newcommand{\pa}[2]{\frac{\partial #1}{\partial #2}}
\newcommand{\paf}[2]{\partial #1 / \partial #2}
\newcommand{\llangle}{\left \langle}
\newcommand{\rrangle}{\right \rangle}

\newcommand{\Wt}{\mathcal{W}}
\newcommand{\etab}{\bar{\eta}}
\newcommand{\gammab}{\bar{\gamma}}

\begin{document}



\title{Momentum-Exchange Current Drive by Electrostatic Waves in an Unmagnetized Collisionless Plasma}

\author{Ian E. Ochs and Nathaniel J. Fisch}
\affiliation{Department of Astrophysical Sciences, Princeton University, Princeton, New Jersey 08540, USA}


\date{\today}

\begin{abstract}
For an electrostatic wave interacting with a single species in a collisionless plasma,  momentum conservation implies current conservation. 
However, when multiple species interact with the wave, they can exchange momentum, leading to current drive. 
A simple,  general formula for this driven current is derived.
As examples, we show how currents can be driven for Langmuir waves in electron-positron-ion plasmas, and for ion-acoustic waves in electron-ion plasmas.

\end{abstract}


\maketitle





\textbf{Introduction:} 
There are a variety of mechanisms through which  electrical current might be driven by waves in plasma \cite{fisch1987theory}.   
A major application area for these mechanisms is the tokamak, which requires steady state plasma currents for confinement, and has consequently dominated the literature on wave-mediated current drive.
%
%

Purely electrostatic waves at first appear to be a strong candidate for current drive.
Such waves interact with particles traveling near the phase velocity, accelerating particles slightly slower than the wave, and decelerating particles faster than the wave.
Since distribution functions usually decrease with energy, the net effect is an acceleration of the resonant particles, driving a current.
However, as shown in the textbook example of {\it plasma quasilinear theory}, which self-consistently describes wave-particle interactions, this is not the case \cite{kralltrivelpiece,bellan2008fundamentals,stix1992waves}. 
While resonant electrons gain momentum from the wave, the non-resonant bulk distribution shifts in the opposite direction, so as to conserve electron momentum.
This cancellation occurs because the electrostatic fields in the plane wave carry no momentum. 

There are several ways to drive current in spite of this constraint.
Most straightforward perhaps is to employ a \emph{quasi}-electrostatic wave, with a small electromagnetic component that creates a momentum flux \cite{stix1992waves, wort1971peristaltic, swanson2012plasma}.
Such waves are particularly relevant to steady-state boundary-value problems, such as lower hybrid current drive from wave antennae in tokamaks \cite{fisch1978currentDrive}, although the distinction between these and purely electrostatic waves is often ignored.
Here, we instead focus on purely electrostatic initial value problems in isolated systems, which are likely to be more relevant in astrophysical settings and well-insulated laboratory devices.

Because of the lack of momentum in a purely electrostatic planar field, a necessary (but not sufficient) condition for electrical current to be generated from a purely electrostatic wave is momentum exchange between multiple species.
Such momentum exchange can be provided by collisions.
For instance, for waves with high phase velocities, because the resonant current is driven in the tail electrons, which are much less collisional than the thermal bulk electrons, the resonant  current will be longer-lived than the nonresonant current.
Thus, a net current is produced on collisional timescales \cite{fisch1978currentDrive,fisch1980creating,stix1992waves,bellan2008fundamentals}.

However, purely electrostatic waves can drive current even absent collisions.  
If a wave interacts strongly with multiple species, it can mediate momentum exchange between the species even in the absence of collisions. 
For species with different charge-to-mass ratios, the conservation of momentum will not imply conservation of current, and net current can be driven.
Such momentum exchange processes have been explored briefly in magnetized plasmas \cite{kato1980electrostatic} and laser-accelerated plasmas \cite{manheimer1977iawCurrent}, but overall received little attention.

Here, we aim to elucidate this current drive mechanism by considering the simple case of an unmagnetized plasma, which is to our knowledge absent in the literature.
Because our approach clearly distinguishes contributions from resonant and nonresonant particles, we can calculate for the first time the growth rate and saturation levels of the resulting currents.

To calculate the driven current, we follow standard plasma quasilinear theory \cite{kralltrivelpiece}, deriving a succinct expression for collisionless current generation via electrostatic waves in multispecies plasma.
We then show how this expression leads to simple calculations of the current growth.
As a first example, we consider Langmuir waves in electron-positron-ion plasmas, which are being produced at increasing densities in laboratory settings \cite{hergenhahn2018apex, sunn2019new, chen2015scaling, chen2018target}.
As a second, we consider ion-acoustic waves in more typical electron-ion plasmas.
This latter example is particularly interesting since there is no separation of collision timescales for resonant and nonresonant particles, so without the wave-mediated momentum exchange no current could be driven.

\textbf{Quasilinear Theory:}
Following standard treatments of quasilinear (QL) theory for 1D  electrostatic waves \cite{kralltrivelpiece}, we solve to first order the Vlasov-Poisson system: 
\begin{align}
	\pa{f_{s}}{t} + v \pa{f_{s}}{x} + \frac{q_s}{m_s} E \pa{f_s}{v} &= 0\\
	\pa{E}{x} - \sum_s 4 \pi q_s n_s \int dv f_s &= 0. 
\end{align}
Here, $q_s$, $m_s$, and $n_s$ are the charge, mass, and zeroth-order density of species $s$, and $f_{s0}$ is the $0$th-order phase-space distribution function normalized to one.
%
Dividing the real and imaginary components of the wave frequency out as $\omega = \omega_r + i \omega_i$, we express the linear dispersion relation derived from this system for $|\omega_i| \ll |\omega_r|$ as:  
\begin{align}
0&=1 + \sum_s D_{r,s} \label{eq:realDisp}\\
0&= \sum_s  \lp i \omega_i \pa{D_{r,s}}{\omega_r} + D_{i,s} \rp \label{eq:imagDisp}.
\end{align}
Here, we have defined the real and imaginary (at real $\omega$) dispersion components associated with each species:
\begin{align}
D_{r,s} &\equiv  -\frac{\omega_{ps}^2}{k^2} PV \int dv \frac{\paf{f_{s0}}{v}}{v-\omega_r/k} \label{eq:Dr1}
\end{align}
\begin{align}
D_{i,s} &\equiv -\pi \frac{\omega_{ps}^2}{k^2} \pa{f_{s0}}{v}|_{\omega_r/k} \label{eq:Di1},
\end{align}
and $\omega_{ps}$ is the plasma frequency of species $s$.
Solving Eq.~(\ref{eq:realDisp}) gives the real component of the frequency $\omega_r$ for a given wavenumber $k$, and solving Eq.~(\ref{eq:imagDisp}) with this $\omega_r$ and $k$ then gives the associated imaginary component of the freqency $\omega_i$.

There are two primary forms of energy associated with the wave. First, there is the electrostatic energy density $W$ associated with the wave electric field $E$:
\begin{align}
	W &\equiv \llangle \frac{E^2}{8\pi}\rrangle = \frac{E_0^2}{16\pi}.
\end{align}
Second, there is the total wave energy density $\Wt$ \cite{stix1992waves,dodin2012axiomatic}, which also incorporates the oscillating kinetic energy:
\begin{align}
	\Wt &= W \left[ \omega_r \pa{}{\omega_r} \lp  \sum_s D_{r,s} \rp \right]. \label{eq:waveVsElectrostatic}
\end{align}

From this linear theory, we derive the quasilinear theory by averaging the Vlasov equation over space and neglecting nonlinear interactions in the evolution of $f_{s1}$.
Thus, the zeroth-order (space-averaged) distribution function evolves to lowest order as:
\begin{gather}
\pa{f_{s0}}{t} = \pa{}{v} \left[\lp \frac{2 \omega_{ps}^2}{m_s n_s} \int_\mathcal{L} \frac{w(k)}{i(kv - \omega)} d k \rp \pa{}{v} f_{s0} \right] \label{eq:QLeqn}\\
\pa{W_k(k)}{t} = 2 \omega_{i}(k,t) W_k(k). \label{eq:dWdt}
\end{gather}
Here, $\mathcal{L}$ denotes the Landau contour, which passes under the poles, and $w(k)$ is the electrostatic energy density stored in the mode $k$, related to $W$ by:
\begin{align}
W &= \int dk \, w(k)  = \frac{1}{V} \int \frac{dk}{2\pi} \frac{E_{k} E_{-k}}{8\pi} ,
\end{align}
where $V$ is the (1D) volume of the wave region.

The evolution of the momentum density $p_s$ and kinetic energy density $K_s$ of each species $s$ are given by multiplying by mass and taking the first and second velocity moments of Eq.~(\ref{eq:QLeqn}).
This integration is easy if we consider a narrow spectrum of waves near $k$ with total energy $W$, such that $w(k') = \frac{W}{2} (\delta(k'-k) + \delta(k'+k))$.
Because Eq.~(\ref{eq:QLeqn}) simply averages the responses to different wavenumbers, this approach determines the characteristic plasma response.
Using the fact that the dispersion relation Eqs.~(\ref{eq:Dr1}-\ref{eq:Di1}) imply $\omega(k,t)=-\omega^*(-k,t)$, and exploiting $|\omega_i/\omega_r| \ll 1$ and the Plemelj formula, we find:
\begin{align}
\frac{dp_s}{dt} &=  2  W k  \left[ \omega_i \pa{}{\omega_r} D_{r,s}+ D_{i,s} \right]\\
\frac{dK_s}{dt} &= \frac{\omega_r}{k} \frac{dp_s}{dt} + 2\omega_i W D_{r,s} .
\end{align}
From these simple equations, it quickly follows from the dispersion relation Eqs.~(\ref{eq:Dr1}-\ref{eq:Di1}) and the electrostatic energy evolution Equation~(\ref{eq:dWdt}) that the total momentum and energy (kinetic + electrostatic) are conserved in the system.
When considering a single species interacting with the wave, the conservation of the momentum implies conservation of the current.

\textbf{Current Drive:} 
In contrast, consider a situation in which multiple species interact with the wave.
In this case, using Eq.~(\ref{eq:waveVsElectrostatic}), we can write the current as:
\begin{align}
	\frac{dj}{dt} &= \sum_s \frac{q_s}{m_s} \frac{dp_s}{dt} = 2  \Wt \frac{k}{\omega_r} \sum_s  \frac{q_s}{m_s} \biggl[   \etab_{s}\omega_i -\omega_{i,s}  \biggr], \label{eq:djdtFundamental}
\end{align}
where we have defined the species damping $\omega_{i,s}$:
\begin{align}
	\omega_{i,s} \equiv -\frac{D_{i,s}}{\sum_{s'} \paf{D_{r,s'}}{\omega_r} },
\end{align}
and the species nonresonant response coefficient $\etab_s$:
\begin{align}
	\etab_{s} \equiv \frac{\paf{D_{r,s}}{\omega_r} }{\sum_{s'} \paf{D_{r,s'}}{\omega_r} }. \label{eq:etab}
\end{align}
Here, $\etab_{s}$ provides a relative measure of how strongly the wave pushes on the nonresonant particles of each species, since $\sum_s \etab_s = 1$.

Assuming that the wave is not at marginal stability (i.e. $d\Wt/dt \neq 0$), we can use Eq.~(\ref{eq:dWdt}) to rewrite the current drive in a very symmetric way:
\begin{align}
\frac{dj}{dt} &=-\frac{d\Wt}{dt} \frac{k}{\omega_r}   \sum_s \frac{q_s}{m_s} \lp   \gammab_s  -   \etab_{s} \rp  \label{eq:djdtdWdt},
\end{align}
where $\gammab_s \equiv \omega_{i,s}/\omega_i$ is a measure of the relative resonant response of each species, with $\sum_s \gammab_s = 1$.

Eq.~(\ref{eq:djdtdWdt}) has a simple physical interpretation.
Consider the case of a light resonant species $l$ and a heavy nonresonant species $h$. 
Assume all of the resonant momentum goes into $l$ ($\bar{\gamma}_l \gg \bar{\gamma}_h$), all of the nonresonant momentum goes into $h$ ($\etab_h \gg \etab_l$), and only current in $l$ contributes significantly ($q_h/m_h \ll q_l/m_l$).
Then, Eq.~(\ref{eq:djdtdWdt}) becomes:
\begin{align}
\frac{dj}{dt} = -\frac{q_l}{m_l} \frac{k}{\omega_r}  \frac{d\Wt}{dt} = \frac{q_l}{m_l}  \frac{1}{v_{ph}}  \frac{d K_l}{dt} = \frac{q_l}{m_l} \frac{dp_{res}}{dt},
\end{align}
where in the last equality we used the fact that if we push a particle near resonance, $v_{ph} dp_s/dt = dK_s / dt$.
Thus, Eq.~(\ref{eq:djdtdWdt}) simply generalizes this equation to include the nonresonant reactions of the various species.

For Langmuir oscillations in an electron-ion plasma, $\gammab_i$ is exponentially small and $\etab_i \sim \mathcal{O}(m_e / m_i)$, so the terms in parentheses in Eq.~(\ref{eq:djdtdWdt}) cancel to $\mathcal{O}(m_e / m_i)$.
However, for Langmuir oscillations in more general plasmas, or for more general plasma waves, current can be driven even for the collisionless electrostatic plasma.

\textbf{Electron-positron-ion plasmas:}
Langmuir waves occur in the frequency range $\omega_r \gg k v_{ths} \; \forall s$.
Asymptotically expanding our integrals for each species yields:
\begin{align}
D_{r,s} &\approx - \frac{\omega_{ps}^2}{\omega_r^2} \label{eq:langDrs}\\
D_{i,s} &\approx -\pi \frac{\omega_{ps}^2}{k^2} \pa{f_{s0}}{v}|_{\omega_r/k} \label{eq:langDis}.
\end{align}

Thus, the nonresonant response Eq.~(\ref{eq:etab}) becomes:
\begin{align}
\bar{\eta}_s &= \frac{n_s q_s^2 /m_s}{\sum_{s'} n_{s'} q_{s'}^2/m_{s'}},
\end{align}
Consider a plasma composed of electrons $e$, ions $i$, and positrons $p$, so that $n_e = Z n_i + n_p$.
Thus $\etab_i \sim \mathcal{O} (m_e/m_i)$, and
\begin{align}
	\etab_e \approx \frac{n_e}{n_e + n_p} \geq \frac{1}{2}; \qquad \etab_p \approx 1-\etab_e.
\end{align}

Thus, from Eq.~(\ref{eq:djdtdWdt}):
\begin{align}
\frac{dj}{dt} &=-\frac{d\Wt}{dt}\frac{k}{\omega_r} \frac{e}{m_e} \left[\lp  \gammab_p - \gammab_e \rp + \lp 2 \etab_e - 1\rp \right]. \label{eq:djdtEPI}
\end{align}
Here, the first term in the brackets is the resonant current drive, and the second term is the nonresonant current drive.
In a pure $e-p$ plasma, the nonresonant currents would cancel, and only the resonant current drive would occur. 
Then, differences in the tail distribution between electrons and positrons can drive resonant current.

In an electron-ion-positron plasma, the imbalance of electrons and positrons can result in currents in two ways.
First, if the pair plasma has a much higher energy-per-particle than the bulk plasma, Langmuir waves on the electron-positron tail will have canceling resonant currents, but the excess of low-energy electrons will result in non-canceling nonresonant currents.
Thus, damping or amplification of Langmuir waves in the pair plasma will drive nonresonant currents in the bulk.

Second, positrons from the pair plasma can annihilate with electrons from the bulk plasma, creating an excess of high-energy electrons.
Then, there will be an imbalance in the kinetic distributions of electrons and positrons, allowing for resonant current drive.

\textbf{Ion Acoustic Waves:}
Now consider ion-acoustic waves (IAWs) in a Maxwellian electron-ion plasma (allowing different temperatures for each species), for which $v_{thi} \ll \omega_r / k \ll v_{the}$.
While the real dispersion Eq.~(\ref{eq:langDrs}) remains valid for ions, for electrons we asymptotically expand in the opposite limit to find (using $f_{s0} \propto e^{-v^2/2v_{ths}^2}, v_{ths} \equiv \sqrt{T_s/m_s}$):
\begin{align}
D_{r,e} &= \frac{1}{k^2 \lambda_{De}^2}\lp 1-\frac{\omega_r^2}{k^2 v_{the}^2} \rp. \label{eq:iawDre}
\end{align}
Here, we retain the second term because it does not vanish upon differentiation by $\omega_r$.

If we also assume the tails are Maxwellian, then solving the dispersion relation Eqs.~(\ref{eq:realDisp}-\ref{eq:imagDisp}) results in the standard IAW frequency and damping rates:
\begin{align} 
\omega_r &= \kappa C_s k \label{eq:iawWr}\\
\omega_{i,i} &= -\sqrt{\frac{\pi}{8}} |\omega_r| \delta_i^{-3} e^{-\delta_i^{-2}/2} \label{eq:iawWii}\\
\omega_{i,e} &= -\sqrt{\frac{\pi}{8}} |\omega_r| \kappa^2 \delta_e,   \label{eq:iawWie}
\end{align}
where $C_s = \sqrt{Z T_e/m_i}$, and $\kappa \equiv (1+k^2 \lambda_{De}^2)^{-1/2} \sim 1$. 
Here, we have defined the small dimensionless parameters associated with the ion acoustic ordering: $\delta_e \equiv \omega_r/kv_{the} = \kappa \sqrt{Z m_e/m_i}$ and $\delta_i \equiv kv_{thi}/\omega_r = \kappa^{-1} \sqrt{T_i/Z T_e}$.
The condition $\delta_i \ll  1$ generalizes the textbook requirement $T_e \gg T_i$ to ions of arbitrary $Z$.

The nonresonant response of each species is found by inserting Eq.~(\ref{eq:langDrs}) for ions and Eq.~(\ref{eq:iawDre}) into Eq.~(\ref{eq:etab}), yielding $\etab_i \approx 1$, and $\etab_e \approx \kappa^2 \delta_e^2 \ll 1$.
Thus, the nonresonant momentum transfer primarily goes to the \emph{ions}.

Meanwhile, for the resonant damping, $\gammab_{e} \lesssim \gammab_i$, i.e. in general there will be more resonant damping on the ions, but $\gammab_e / \gammab_i \gg \delta_e \gg \eta_e$, which can be seen from Eq.~(\ref{eq:iawWie}) and the requirement that $|\omega_i / \omega_r| \ll 1$.

With these basic orderings for our resonant and nonresonant response coefficients, 
we can write Eq.~(\ref{eq:djdtdWdt}) as:
\begin{align}
\frac{dj}{dt} &=\frac{d\Wt}{dt} \frac{k}{\omega_r}  \frac{e}{m_e} \left[ \lp   \gammab_e  -   \etab_e \rp + \delta_e^2 \lp   \gammab_i  - \etab_{i}  \rp\right]\\
&= \frac{d\Wt}{dt} \frac{k}{\omega_r}  \frac{e}{m_e} \gammab_{e} \lp 1 + \mathcal{O}(\delta_e) \rp,
\end{align}
i.e. the resonant electron current dominates all other contributions by a factor of $\delta_e$.

Knowing that the resonant electron current dominates, we can substitute $\omega_{i,e} = \omega_i \etab_e$ and use Eq.~(\ref{eq:dWdt}) to write:
\begin{align}
\frac{dj}{dt} &= 2 \Wt \omega_{i,e} \frac{k}{\omega_r}  \frac{e}{m_e}  \lp 1 + \mathcal{O}(\delta_e) \rp. \label{eq:iawdjdtFinal}
\end{align}
This form of the equation makes it clear that the resonant current direction does not depend on the overall stability of the plasma, only on the sign of the electron damping.

It is critical to note that this resonant current would not appear even in a collisional analysis if the nonresonant response was not primarily in the ions, since there is no collisional timescale separation between the resonant and nonresonant electrons.
The wave-mediated momentum exchange is thus fundamentally required for this current drive mechanism.

\begin{figure}[t]
	\center
	\includegraphics[width=\linewidth]{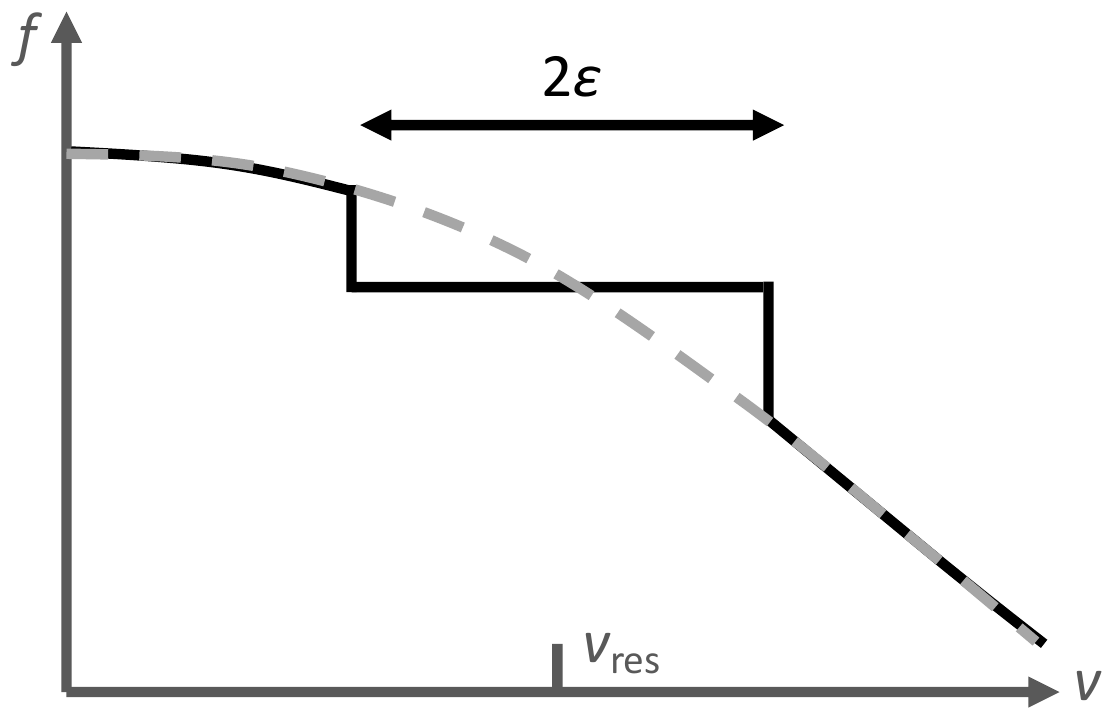}
	\caption{Saturation of the quasilinear resonant current.
		The initial distribution $f_0(v)$ (gray dashed) flattens out in a region of width $2 \epsilon v_\text{res}$ around the resonance distribution, resulting in the final distribution (black solid).
	}
	\label{fig:plateau}
\end{figure}

\textbf{Kinetic Saturation:}
One of the advantages of our approach, which distinguishes resonant and nonresonant currents, is that it allows us to estimate the saturation level of the current beyond the linear regime.
As the wave damps, each resonant species' distribution function $f_{s0}(v)$ around the resonance will flatten, and eventually the damping will stop when $f_{s}(v)$ flattens completely in some region $(1-\epsilon)v_\text{res} < v < (1+\epsilon)v_\text{res}$ (Fig.~\ref{fig:plateau}).
If all the current is driven resonantly, as for the ion-acoustic wave, then the saturated current is simply the difference in current between this final flattened distribution and the initial distribution.

For ion-acoustic waves, the electron distribution function $f_{e0}(v)$ flattens in a resonance region of width $\epsilon C_s$ around $v = C_s$.
This broadening can be provided in two ways. 
First, if there is a spectrum of waves present with different values of $k$, then from Eq.~(\ref{eq:iawWr}) and the definition of $\kappa$, $\epsilon = k_{\max}^2 \lambda_{De}^2/2$.
Second, a finite-amplitude wave will nonlinearly trap electrons within $\epsilon = \sqrt{8 e E_0/m_e k C_s^2}$.

After flattening, the distribution function in the resonance region will everywhere assume its initial average value, i.e. 
\begin{equation}
	f_{ef} = \langle f_{e0} \rangle_{res} = \frac{1}{\epsilon C_s}\int_{(1-\epsilon) C_s}^{(1+\epsilon) C_s} dv f_{e0}.
\end{equation}
If we keep only the lowest-order terms for a Maxwellian distribution, we find to lowest order in $\epsilon$:
\begin{align}
\Delta j_\text{max} &= q_e n_e \int dv \, v (f_{ef} - f_{e0})\\
&\approx \frac{\epsilon^3}{6\sqrt{2\pi}} \lp \frac{C_s}{v_{the}} \rp^3 (q_e n_e C_s). \label{eq:iawSaturation}
\end{align}
Remarkably, the final term in parentheses represents quite a large current, i.e. all the electrons flowing at the sound speed, so that even with the many small terms in front of it, the current can still be quite large.

\textbf{Conclusions:}
We derived for the first time a simple, general expression for the current drive generated by an electrostatic wave in an unmagnetized, collisionless plasma.
We  applied this expression to show how current can be generated by Langmuir waves in electron-positron-ion plasmas, and ion-acoustic waves in electron-ion plasmas.
Because our approach distinguishes resonant and nonresonant currents, we were able to calculate the saturated collisionless current for the first time.

The wave-mediated momentum exchange we derived is the simplest example of a largely neglected current drive effect \cite{manheimer1977iawCurrent,kato1980electrostatic}, which can operate in systems with neither a collisional timescale separation between resonant and nonresonant particles, nor a magnetic component to the wave.

\textbf{Acknowledgments:}
We would like to thank E.J. Kolmes and M.E. Mlodik for helpful discussions.
This work was supported by grants  DOE DE-SC0016072 and   DOE NNSA DE-NA0003871.
One author (IEO) also acknowledges the support of the DOE Computational Science Graduate Fellowship (DOE grant number DE-FG02-97ER25308).


%

\clearpage\newpage

\end{document}